\documentclass{ws-procs975x65}

\usepackage{amsmath}
\usepackage{amsfonts}
\usepackage{amssymb}
\usepackage{dsfont}
\usepackage{latexsym}
\begin{document}

\title{Evolutionary Reformulation of Quantum Gravity}

\author{Giovanni Montani$^{*\dag}$}

\address{$^*$ICRA---International Center for Relativistic Astrophysics\\
Dipartimento di Fisica (G9), Universit\`a  di Roma, ``La Sapienza",\\
Piazzale Aldo Moro 5, 00185 Rome, Italy.}

\address{$^\dag$ENEA-C.R. Frascati (U.T.S. Fusione),\\ via Enrico Fermi 45, 00044 Frascati, Rome, Italy.}

\address{E-mail: montani@icra.it}

\begin{abstract}
We present a critical analysis of the Canonical approach
to quantum
gravity, which relies on the ambiguity of implementing
a space-time
slicing on the quantum level. We emphasize that
such a splitting
procedure is consistent only if a real matter
fluid is involved in the
dynamics.
\end{abstract}

\bodymatter

\vspace{0.5cm}

General Relativity is a background independent theory
which identify the gravitational interaction into the metric
properties of the space-time and this peculiar nature makes very
subtle even simple questions about its quantization. To deal with a
canonical method for the fields dynamics necessarily involves the
notion of a physical time variable, whose conjugate momentum
fixes the Hamiltonian function.
Already in the context of classical General Relativity,
the task of recovering a physical clock 
acquires non-trivial character, depending on the local properties of
the space-time. However, a well grounded algorithm devoted to this end was
settled down by Arnowitt-Deser-Misner (ADM) in \cite{ADM62}
It consists of a space-time slicing based on a one parameter family
of spacelike hypersurfaces $\Sigma ^3_t$, defined via the parametric
representation $t^{\mu } = t^{\mu }(t,\; x^i)$
($\mu = 0,1,2,3$ and $i = 1,2,3$). 
In what follows, we denote the set of coordinates
$\{ t,\; x^i\}$
by $x^{\bar{\mu}}$, in order to emphasize that the slicing
procedure can be recast as a 4-diffeomorphism, i.e.
$ds^2 = g_{\mu \nu }(t^{\rho})dt^{\mu }dt^{\nu } =
\bar{g}_{\bar{\mu }\bar{\nu }}(x^{\bar{\rho}})
dx^{\bar{\mu }}dx^{\bar{\nu }}$.
The main issue of adopting the coordinates $x^{\bar{\mu }}$ is
that they allow to separate the 4-metric tensor into six
evolutionary components, which determine the induced 3-metric
tensor $h_{ij}$ of the hypersurfaces and four Lagrangian
multipliers, corresponding to the lapse function $N$ and to the
shift vector $N^i$. These non-evolutionary variables have a
precise geometrical meaning, given by the relation
$\partial _tt^{\mu } = Nn^{\mu } + N^i\partial _it^{\mu }$
(where $g_{\mu \nu } n^{\mu }n^{\nu } = 1$ and 
$g_{\mu \nu } n^{\mu }\partial _it^{\nu } = 0$),
$n^{\mu }(t^{\rho })$ denoting the orthonormal vector to the
family $\Sigma ^3_t$. The classical dynamics of the
3-metric $h_{ij}$ is governed, in vacuum, by the following set of
equations

\begin{eqnarray}
\label{c}
G_{\mu \nu}n^{\mu }n^{\nu } = - \frac{kH}{2\sqrt{h}}=0\\
G_{\mu \nu}n^{\mu }\partial _it^{\nu } = \frac{kH_i}{2\sqrt{h}}=0\\
G_{\mu \nu}\partial _it^{\mu }\partial _jt^{\nu } =
G_{ij} = 0
\, ,
\end{eqnarray}

$h$ being the 3-metric determinant
and $G_{\mu \nu}$ the Einstein tensor
($H$ and $H_i$ are called the super-Hamiltonian and the
super-momentum respectively). The first two lines above
correspond to constraints for the initial values problem and they
play a crucial role in the canonical quantization of the system,
while the last line fixes the evolution of the 3-metric and it is
lost on the quantum level. As a consequence, the canonical quantum
dynamics of the gravitational field is characterized by the
so-called \emph{frozen formalism} \cite{K81}.In fact the dynamics
of a generic state
$\mid g_{\bar{\mu }\bar{\nu }}\rangle =
\mid h_{ij}, \; N, \; N^i\rangle $ is provided by the requests

\begin{eqnarray}
\hat{p}_N\mid h_{ij}, \; N, \; N^i\rangle = 0 , \;
\hat{p}_{N^i}\mid h_{ij}, \; N, \; N^i\rangle = 0\\
\hat{H}\mid h_{ij}, \; N, \; N^i\rangle = 0, \;
\hat{H}_i \mid h_{ij}, \; N, \; N^i\rangle = 0
\, ,
\end{eqnarray}

$\hat{p}_N$ and $p_{N^i}$ being the momenta operators
associated to $N$ and $N^i$ respectively. The four operator
constraints listed above are the quantum translation of the
diffeomorphisms invariance of the theory
and they can be summarized by the Wheeler-DeWitt equation
\cite{D67} 
$\hat{H}\mid \{ h_{ij}\}\rangle = 0$, where by $\{ h_{ij}\}$  
we denote a class of 3-geometries. The
\emph{frozen formalism} consists of the independence
that the states acquires from $N$ and $N^i$, i.e.
the quantum picture is the same on each spacelike
hypersurfaces.\\
This non evolutionary character of the Wheeler-DeWitt
approach is striking in contrast with the Einstein
equations which predict a 3-metric field evolving over the
slicing. In what follows, we argue that this
absence of a proper time in canonical quantum gravity
is connected to the inconsistency of the 3+1-splitting
referred to a quantum (vacuum) space-time.
As issue of this criticism, we outline a
\emph{time-matter} dualism
and provide an evolutionary re-formulation of the
canonical paradigm for the gravitational field quantization.

Let us assume to have solved the quantum gravity problem
in the framework of generic coordinates $t^{\mu }$, having
determined a complete set of orthonormal states
$\mid g_{\mu \nu }\rangle _{\alpha }$ on which a given
configuration  $\mid g_{\mu \nu }\rangle$ can be decomposed as
$\mid g_{\mu \nu }\rangle
= \sum_{\alpha }c_{\alpha }\mid g_{\mu \nu }\rangle _{\alpha }$.
Now, assigned a 4-vector $n^{\mu }$, its norm
$n\equiv g_{\mu \nu}n^{\mu }n^{\nu }$ 
(and therefore its timelike character too) can be established 
only in the sense of expectation values on the state
$\mid g_{\mu \nu }\rangle$, having the form
$\langle n \rangle = \sum_{\alpha }
c_{\alpha }\langle n\rangle _{\alpha }
\equiv s (t^{\mu })$. This field $s$ is clearly a
random scalar one, whose dynamics is induced by the quantum
behavior of the 4-metric $g_{\mu \nu }$. By the diffeomorphism
invariance, we deal with a scalar field
$s(t^{\rho }(x^{\bar{\mu}})) = s(t, x^i)$ on the slicing picture too.
Analogous considerations hold for the quantity
$n_i \equiv g_{\mu \nu }n^{\mu }\partial _it^{\nu }$
(which states the timelike nature of $n^{\mu }$ in the coordinates
$x^{\bar{\mu }}$) and leads to conclude that its expectation value  
$\langle n_i\rangle = \sum _{\alpha }c_{\alpha }
\langle n_i\rangle _{\alpha } \equiv s_i (t^{\mu })$
define in turn a random vector
$s_i(t^{\rho }(x^{\bar{\mu }})) = s_i(t, x^i)$
living on the 3-hypersurfaces $\Sigma ^3_t$.
The outcoming of these four degrees of freedom
$\{ s,\: s_i\}$ indicates that, for a quantum space-time,
the slicing picture preserves the number of evolutionary variables,
because we pass from $g_{\mu \nu }$ in the system
$t^{\mu }$ to $\{h_{ij},\; s,\; s_i\}$ in the splitting coordinates
$x^{\bar{\mu }}$. In this respect $N$ and $N^i$ simply give the
components of the vector $n^{\mu }$ in the 3+1-scheme.
Now, the evolutionary behavior of
ten variables 
(right the number of 4-metric components)
implies that the super-Hamiltonian and the super-momentum
constraints (\ref{c}) are violated in the sense of expectation
values, so that
$\langle k\hat{H}/2\sqrt{h}\rangle = \varepsilon$ and 
$\langle k\hat{H}_i/2\sqrt{h}\rangle = q_i$. Here $\varepsilon$
and $q_i$ denote a 3-scalar and a 3-vector field respectively.
Their presence comes out because of the equation $G_{ij} = 0$,
which classically ensure the existence of constraints,
are lost on a quantum level. By other words, if we quantize
the gravitational field before the slicing procedure is performed,
then the quantum translation of the 3+1-picture can no longer be
recovered and the \emph{frozen formalism} is overcome. 

The physical issue of the analysis above, leads to a
\emph{time-matter dualism} within the context of an evolutionary
quantum gravity. In fact, the following two statements take place
on the quantum and classical level respectively.\\
i) The non-vanishing behavior of the super-Hamiltonian and
the super-momentum expectation values implies that the corresponding
operators do not annihilate the states of the theory
(like in the Wheeler-DeWitt approach) and therefore we have to deal
with a schr\"odinger quantum dynamics of the gravitational
field \cite{M02}. More precisely, in the coordinates $x^{\bar{\mu }}$
the state acquires a dependence on the label time, i.e.
it reads $\mid t,\; h_{ij}\rangle$ and it obeys the Schr\"odinger
equation 

\begin{equation}
i\hbar \partial _t
\mid t,\; h_{ij}\rangle =
\hat{\mathcal{H}}
\mid t,\; h_{ij}\rangle \equiv
\int _{\Sigma ^3_t}d^3x\left\{
N\hat{H} + N^i\hat{H}_i\right\}
\mid t,\; h_{ij}\rangle
\, ;
\end{equation}

this equation provides the time evolution of the
3-metric states along the slicing and, once fixed
the proper operator ordering to deal with an
Hermitian Hamiltonian, then a standard procedure
defines the Hilbert space.\\
ii) The classical WKB limit for
$\hbar \rightarrow 0$ maps the Schr\"odinger
dynamics above into the relaxed Hamiltonian
constraints, which contain $\varepsilon$
and $q_i$ 
\cite{M02}. By using the relations (\ref{c}),
the classical limit is recognized to be General
Relativity in presence of an Eckart fluid \cite{E40},
i. e.

\begin{equation}
G_{\mu \nu } =  k\left(
-\varepsilon n_{\mu }n_{\nu } +
n_{\mu }q_{\nu } +
q_{\mu }n_{\nu }\right) , \quad
q_{\mu } \equiv q_ih^{ij}\partial _jt^{\mu }
\, ;
\end{equation}

above, $h^{ij}$ denotes the inverse
3-metric and the 4-vector $q_{\mu }$ has the physical
meaning of heat conductivity. Here $n^{\mu }$ plays
the role of 4-velocity, according to the request of a
physical slicing which preserves the light cone
on a quantum level too.\\ 
It is relevant to stress that the energy density of the
Eckart fluid is positive in correspondence to the negative
part of the super-Hamiltonian spectrum. Therefore,
showing that such a region is predicted by the quantum dynamics
acquires here a key role.

Having this idea in mind, we
adopt more convenient variables to express the
3-metric tensor, i.e.

\begin{equation}
h_{ij} \equiv \eta ^{4/3}u_{ij}
\, ,
\label{detvar}
\end{equation}

with $\eta \equiv h^{1/4}$ and $det u_{ij} = 1$.\\
Expressed via these variables, the
super-Hamiltonian reads

\begin{equation}
H = -\frac{3}{16}c^2 kp_{\eta }^2 +
\frac{2c^2 k}{\eta ^2}u_{ik}u_{jl}p^{ij}p^{kl}
- \frac{1}{2k}\eta ^{2/3}V(u_{ij},\; \nabla \eta ,\; \nabla u_{ij})
\label{newh}
\, ,
\end{equation}

where $p_{\eta }$ and $p^{ij}$ denote the conjugate momenta to
$\eta $ and $u_{ij}$ respectively, while  
the potential term $V$ comes from the 3-Ricci scalar and
$\nabla$ refers to first and second order spatial gradients.

In this picture, 
the eigenvalue of the super-Hamiltonian operator takes the
explicit form

\begin{eqnarray}
\label{newh}
\hat{H}\chi _{\mathcal{E}} =
\left\{ 
\frac{3}{128\hbar ck^2}\frac{\delta ^2 }{\delta \eta ^2} -
\frac{1}{4\hbar ck^2\eta ^2}\Delta _{u}                          
- \frac{1}{2k}\eta ^{2/3}V(u_{ij},\; \nabla \eta ,\; \nabla u_{ij})
\right\}\chi _{\mathcal{E}}  = 
\mathcal{E}\chi _{\mathcal{E}}\\
\Delta _{u} \equiv
\frac{\delta }{\delta u_{ij}}
u_{ik}u_{jl}
\frac{\delta }{\delta u_{kl}}
\, .
\end{eqnarray}

From a qualitative point of view, the existence of solutions
for the system (\ref{newh}) with negative values of
$\mathcal{E}$ can be inferred from its Klein-Gordon-like
structure. However, a more quantitative analysis 
is allowed by taking the limit
$\eta  \rightarrow 0$, where the system 
(\ref{newh}) admits an asymptotic solution.
In fact, in this limit, the potential term is
drastically suppressed with respect to the $\Delta _{u}$ one
and the dynamics of different spatial points decouples,
so reducing the quantization scheme to the
local minisuperspace approach. 
It is easy to see that such approximate
dynamics admits, point by point in space, the solution

\begin{equation}
\chi _{\mathcal{E}} =
\iota _{\mathcal{E}}(\eta , p)G_{p^2}(u_{ij})
\, ,
\label{iog}
\end{equation}

$\iota $ and $G_{p^2}$ satisfying the two equations
respectively

\begin{eqnarray}
\label{twoeig1}
\left\{ 
\frac{1}{\hbar ck^2}\frac{\delta ^2}{\delta \eta ^2} +
\frac{32p^2}{\hbar ck^2\eta ^2}                                    
\right\}\iota _{\mathcal{E}} = 
\mathcal{E}\iota _{\mathcal{E}}\\
\Delta _{u} G_{p^2} = - p^2G_{p^2}
\, .
\end{eqnarray}

As far as we take $\iota = \sqrt{\eta } \theta (\eta )$ and
we consider the negative part of the spectrum
$\mathcal{E} = -\mid \mathcal{E}\mid$, the function
$\theta $ obey the equation

\begin{eqnarray}
\label{twoeig2}
\frac{1}{\hbar ck^2}\frac{\delta ^2\theta }{\delta \eta ^2} +
\frac{1}{\hbar ck^2\eta }\frac{\delta \theta }{\delta \eta } +
\left(\mid \mathcal{E}^{\prime }\mid  -
\frac{q^2}{\hbar ck^2\eta ^2}\right) 
\theta = 0 \\                     
q^2 \equiv \frac{1}{4}\left(1 - 128p^2\right)
\, \quad \mathcal{E}^{\prime } \equiv
\hbar ck^2\mathcal{E}
\, .
\end{eqnarray}

Thus, we see that a negative part of the spectrum exists
in correspondence to the solution

\begin{equation}
\theta (\eta ,\; \mathcal{E}^{\prime },\; p) =
AJ_q(\sqrt{\mid \mathcal{E}^{\prime }\mid } \eta ) +
BJ_{-q}(\sqrt{\mid \mathcal{E}^{\prime }\mid }\eta )
\, ,
\label{bessel}
\end{equation}

where $J_{\pm q}$ denote the corresponding
Bessel functions, while $A$ and $B$ are two integration
constants. 

The above analysis states that                     
the Eckart energy density always has
a (quantum) range of positive value,
(associated to the negative portion
of the super-Hamiltonian spectrum) near enough to the
``singular'' point $\eta = 0$.

However, the correspondence between $\varepsilon$ and
$\mathcal{E}/2\eta ^2$ can occur only after the classical 
limit of the spectrum is taken.\\
We conclude this analysis, observing that 
to give a precise
physical meaning to this picture, the following
three points (elsewhere faced) 
have to be addressed.\\ 
i) The existence of a stable ground level of negative
energy has to be inferred or provided by additional conditions.
ii) The spatial gradients
of the dynamical variables 
and therefore the associated super-momentum constraints,
are to be included into the problem and 
treated in a consistent way.
iii) The physical nature of the limit
$\eta \rightarrow 0$ has to be clarified within a
cosmological framework \cite{BM06}.

We conclude by observing that
reliable investigations \cite{BM06,M03}
provide negative components of the super-Hamiltonian spectrum
which are associated to the constraint

\begin{equation}
\mid \mathcal{E} \mid \le \frac{\hbar c}{l_{Pl}^4}
\, ,
\label{conste}
\end{equation}

where $l_{Pl} \equiv \sqrt{\frac{G\hbar}{c^3}}$ denotes the
Planck length.

As shown in \cite{M03}, this range of variation for
the super-Hamiltonian eigenvalue implies, 
when an inflationary scenario is addressed, 
a negligible contribution
to the actual Universe critical parameter.
In fact, estimating the critical
parameter associated to the new matter term, say
$\Omega _{\mathcal{E}}$, we get

\begin{equation}
\Omega _{\mathcal{E}} \le 
\mathcal{O}\left( \frac{10^{-2}l_{Pl}}{R_0} \right) \sim 
\mathcal{O}\left( 10^{-60}\right)
\, ,
\label{crit}
\end{equation}

$R_0\sim \mathcal{O}\left( 10^{28}cm \right)$ being the present
radius of curvature of the Universe.

Therefore, by above, we see that the predictions of an 
evolutionary quantum cosmology phenomenologically overlap
those ones of the Wheeler-DeWitt approach.

\end{document}